# Detection of Aneuploidy with Digital PCR

H. Christina Fan and Stephen R. Quake

Department of Bioengineering, Stanford University and Howard Hughes Medical Institute, Stanford CA 94305

Abstract

The widespread use of genetic testing in high risk pregnancies has created strong interest in rapid and accurate molecular diagnostics for common chromosomal aneuploidies. We show here that digital polymerase chain reaction (dPCR) can be used for accurate measurement of trisomy 21 (Down's Syndrome), the most common human aneuploidy. dPCR is generally applicable to any aneuploidy, does not depend on allelic distribution or gender, and is able to detect signals in the presence of mosaics or contaminating maternal DNA.

Introduction

At present, amiocentesis and chorionic villus sampling are the most routinely employed clinical procedures to obtain fetal genetic materials for prenatal diagnosis. The gold standard analytical technique to analyze such samples is karyotyping via Giemsa banding; while it is accurate and informative, it is also time consuming and the two weeks between sampling and result can create anxiety for the expectant parents. Conventional karyotyping is therefore gradually being supplanted by techniques based on polymerase chain reaction (PCR) or fluorescence in situ hybridization (FISH)[1, 2]. Although these techniques are substantially faster than karyotyping, they have limitations. FISH is labor intensive and requires intact cells. Quantitative fluorescence PCR (QF-PCR) techniques depend on the existence of allelic variation in the population and therefore are not applicable to homozygotic individuals, and furthermore require an electrophoretic separation step after amplification. There have been attempts to use quantitative PCR (Q-PCR) to perform single step amplification and detection, but it is challenging to achieve the required level of quantitation[2].

Digital PCR is a technique to quantify the amount of DNA in a sample by counting amplifications from single molecules. Digital PCR in principle allows a much higher precision of measurement than Q-PCR, and does not suffer from the limitations of other molecular methods. Briefly, the DNA sample is diluted and aliquoted into compartments or wells such that on average there is less than 1 copy per well. The number of wells that give a positive signal after a PCR reaction corresponds to the number of DNA template molecules in the original sample. A version of this technique involving the use of magnetic beads in microemulsions has been used for early detection of colorectal tumors [3, 4] and as a sample preparation technique for DNA sequencing[5,6]. A microfluidic version of digital PCR has been used for single cell gene expression in hematopoietic stem cells[7] and for mapping gene function to organism identity in environmental microbes[8].

Here we demonstrate the use of microfluidic digital PCR for the detection of chromosomal aneuploidy using Down Syndrome (trisomy 21) as an example. A single microfluidic chip enables the execution of 9,180 parallel PCR reactions, each with nanoliter volume. This method of detection is not sensitive to fluctuations in the amplification efficiency, and therefore offers greater precision than Q-PCR techniques that track the exponential growth of the sample. Counting the number of positive reactions derived from individual template molecules allows one to quantitate the starting material with nearly arbitrary precision, thereby enabling direct measurement of aneuploidy with high confidence.

Experimental Section

Human genomic DNA from a normal cell line and a trisomy 21 cell line were purchased from ATCC (VA, USA). Mixtures containing various proportions of the two genomic DNA samples were also prepared. Primers and Taqman probes specific for amyloid on chromosome 21 and GAPDH on chromosome 12 were adapted from Zimmermann *et al* [9] and purchased from IDT. 10 ul of reaction mixture containing 1X iQ Supermix (BioRad), (300 nM) primers, (150 nM) probes, Tween20 (0.1%), and template DNA at the appropriate dilution was loaded to each panel of a 12.765 Digital Array (Fluidigm, CA) microfluidic chip. The microfluidic chip has 12 panels, each containing 765 partitions. 40 cycles of 2-step PCR was performed on an fluorescence imaging thermal cycler system (Fluidigm, CA) according to manufacturer's instructions. A MATLAB program was written to subtract the image of the chip taken before cycling from that taken at cycle 40 for each fluorescent channel. The number of positive wells in each fluorescent channel was counted.

## Results and Discussion

**Microfluidic Digital PCR**

Digital PCR was performed with genomic DNA derived from a normal human cell line, from a trisomy 21 cell line, and mixtures of both with percentages of trisomy 21 DNA varying from approximately 30% to 60%. Figure 1 shows fluorescent images of a Digital Array chip containing amplification products from genomic DNA derived from the normal and trisomy 21 cell lines. Figure 1a is an image showing amplification from amyloid templates, while Figure 1b is an image showing amplification from GAPDH templates. Each white square represents a compartment containing amplified products. On each image, the left five panels are replicates of amplification from normal genomic DNA, while the five panels on the right are replicates of amplification from trisomy 21 genomic DNA. The top two panels are no template controls. Visual inspection of the images clearly distinguishes normal samples from trisomy 21 samples, since the number of positive compartments for amyloid is greater than that for GAPDH in the case of trisomy 21. Counting of positive squares reveals that the ratio of amyloid to GAPDH is approximately 1:1 for the normal case and approximately 3:2 for the case of trisomy 21.

The distinction between trisomy 21 and normal samples is exceptional; one can resolve the data distributions with no ambiguity and a Student's t-test gives an extremely small p-value (on the order of $10^{-13}$).

The predictive power of a single test panel can be estimated by withholding a data point and asking how well the rest of the data can be used to predict whether the withheld point is normal or Down's positive. A simple model that assumes Gaussian distributions shows that individual test results have p-values of better than $10^{-5}$.

**Sensitivity to Maternal Cell Contamination and Mosaicism**

Real world samples are sometimes characterized by fetal mosaicism or contamination with maternal cells. To assess the robustness of the digital PCR method for the detection of fetal aneuploidy in these circumstances, we performed the assay on samples containing mixtures of normal and trisomy 21 genomic DNA in different proportions.

Figure 2 is a plot of the ratio of amyloid to GAPDH copy number for samples containing different percentages of trisomy 21 genomic DNA (i.e. trisomy 21 DNA/(trisomy 21 DNA + normal DNA)*100%). Each point represents the ratio obtained from one panel (i.e. one replicate), corresponding to 765 compartments. Approximately one third of the compartments in a panel contain amplification products. The mean and standard deviation of the ratio from the normal group are calculated and the boundary representing a 95% confidence interval is drawn (mean + 1.96*standard deviation). If this boundary is used to distinguish a normal karyotype from an abnormal one, a single panel of 765 compartments is capable of making such differentiation for samples containing approximately 40% to 100% trisomy 21 DNA. Student's t-test reveals that all abnormal samples have significantly different amyloid to GAPDH ratio from the normal sample ($p<0.05$).

The ultimate sensitivity of digital PCR to resolve fetal genotype in the presence of maternal contamination is determined by the number of reactions performed. The larger number of molecules assayed, the better is the confidence with which one can make the determination of normal or aneuploid. This can be calculated precisely, as described below.

If abnormal DNA constitutes $\varepsilon*100$ % of the total amount of extracted DNA, for every $m$ genomic equivalents of total DNA, there exist $2m$ copies of a gene on a normal chromosome and $m(2+\varepsilon)$ copies of a gene on the abnormal chromosome. The difference of $m\varepsilon$ is in principle detectable, provided that the difference is at some level greater than the sampling noise, which scales as the square root of copy number. The number of compartments required to achieve $k$ standard deviation of significance can then be estimated (see supporting materials for derivation). $k$ is a function of $\varepsilon$ and the number of compartments used. Figure 3 is a plot of $k$ against number of compartments for various percentages of abnormal DNA in a mixture of normal and abnormal DNA, assuming that every three compartments contain one template molecule. In theory, if $k$ is greater than 1.96, one is 95% confident that the sample is not from a normal fetus. Such a confidence interval can be achieved using approximately 4000 compartments for samples containing 10% abnormal DNA. In this manner, one can achieve nearly arbitrary sensitivity simply by increasing the scale of the assay.

Our calculations show that a sample containing roughly $10^3$ genomic equivalents would enable the determination of normal versus abnormal for sample containing only 10% abnormal DNA with 95% confidence. Such amount of material can be obtained readily from conventional amniocentesis or chorionic villus sampling.

Conclusion

We described a novel molecular diagnostic technique to measure chromosomal aneuploidy. The method is based on the concept of digital PCR, which involves the precise quantitation of single template molecules. Although this is only a proof-of-principle study, we believe that digital PCR would be as competitive as, if not better than, the several molecular diagnostic tools currently practised (such as FISH, QF-PCR and Q-PCR), since it is a rapid, simple, yet precise procedure and is able to detect abnormality with the presence of maternal contamination or fetal chromosomal mosaicism. This method may also be applicable to non-invasive prenatal testing, in which a small amount of fetal DNA is found circulating in the maternal bloodstream.

Acknowledgements

This study was supported in part by the Wallace H. Coulter Foundation Translational Partnership Award.

Supporting Materials

For *m* genome equivalents,

Let *Y*=copy number of Chromosome 21 and *X*=copy number of Chromosome 12

The difference between copy numbers of Chr21 and Chr12 = $D = Y-X = m\varepsilon$

Require the difference *D* to be some constant *k* times the sampling noise $\sqrt{Y}$

$$D = m\varepsilon = k\sigma = k\sqrt{Y}$$

$$m \approx \frac{Y}{2+\varepsilon}$$

$$k = \frac{\varepsilon\sqrt{Y}}{2+\varepsilon}$$

If 1/3 of the panel is used (i.e., 1 positive compartment in every 3 compartments):

*N*=number of compartments

*Y\**=N/3

The response characteristic of digital PCR is developed and presented in Warren *et al*[7].

$$Y = \frac{\log(1-\frac{Y^*}{3})}{\log(1-\frac{1}{N})} = \frac{\log(\frac{2}{3})}{\log(1-\frac{1}{N})}$$

Substitute the expression of *Y* into the expression of *k* to obtain the number of compartments *N* (or PCR reactions required) in order to achieve *k* standard deviations of significance.

Figure Captions

**Figure 1. Fluorescent images of a 12-panel Digital Array chip containing amplified products from normal human genomic DNA (left five panels) and trisomy 21 DNA (right five panels). Top two panels are no template controls. Each white square represents a compartment containing amplified products. Left (Figure 1a): amyloid amplified products (chromosome 21). Right (Figure 1b): GAPDH amplified products (chromosome 12).**

**Figure 2. Plot of the ratio of the amyloid to GAPDH copy number for samples containing different percentages of trisomy 21 genomic DNA (i.e. trisomy 21 DNA/(trisomy 21 DNA + normal DNA)*100%). Each point represents the ratio of one panel, corresponding to 765 compartments. Dash line represents the 95% confidence interval of the normal group.**

**Figure 3. Estimation of the required number of compartments in order to achieve certain level of confidence for various percentages of trisomy 21 DNA (i.e. trisomy 21 DNA/(trisomy 21 DNA + normal DNA)*100%).**

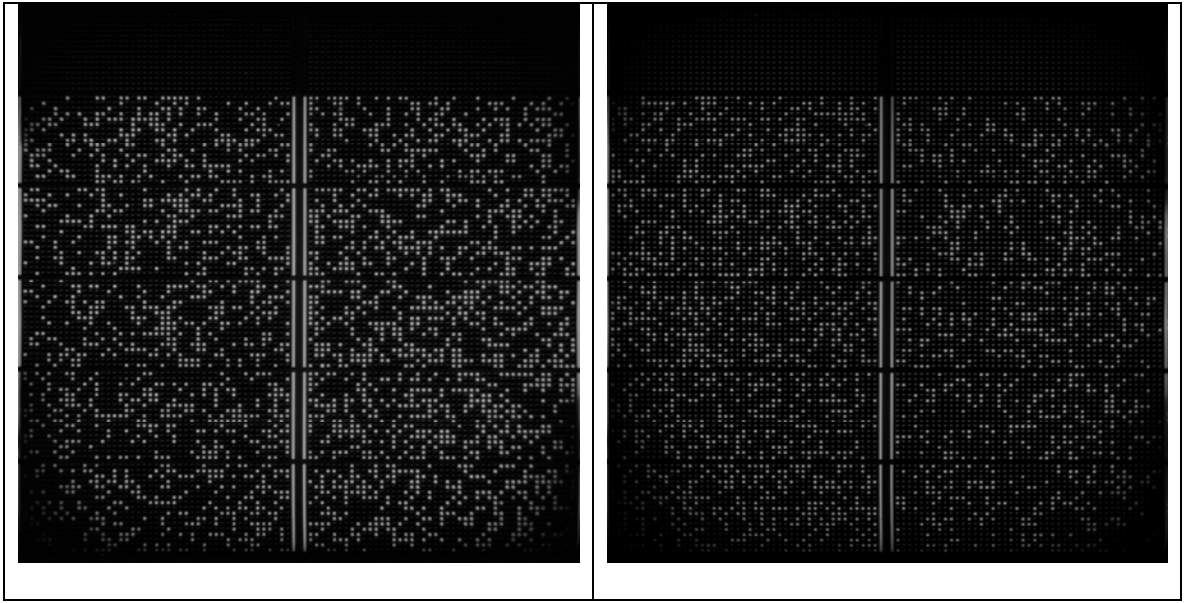

Figure 1

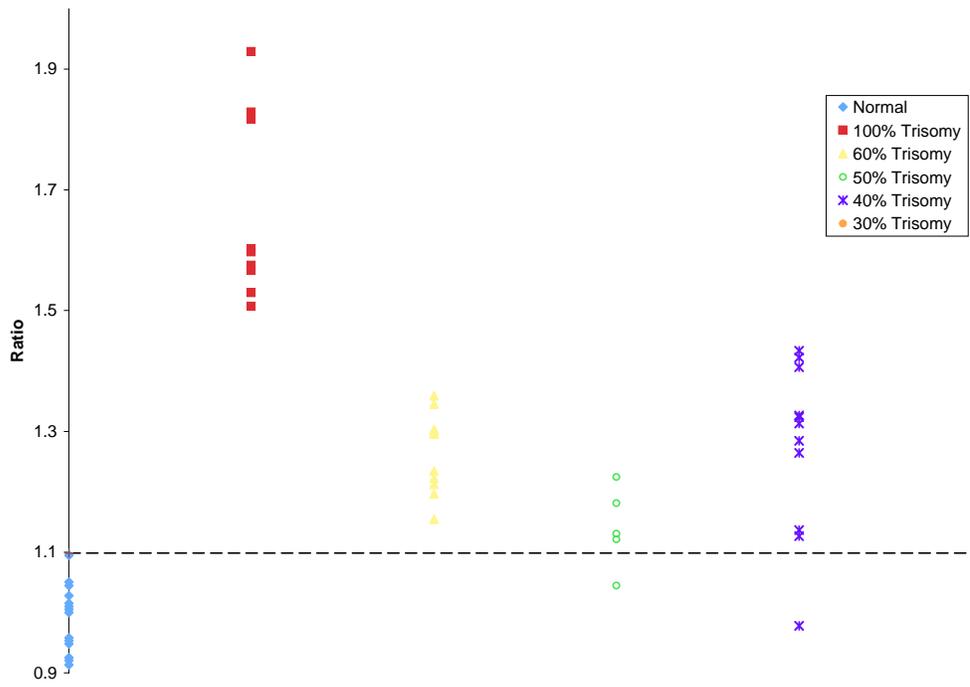

Figure 2

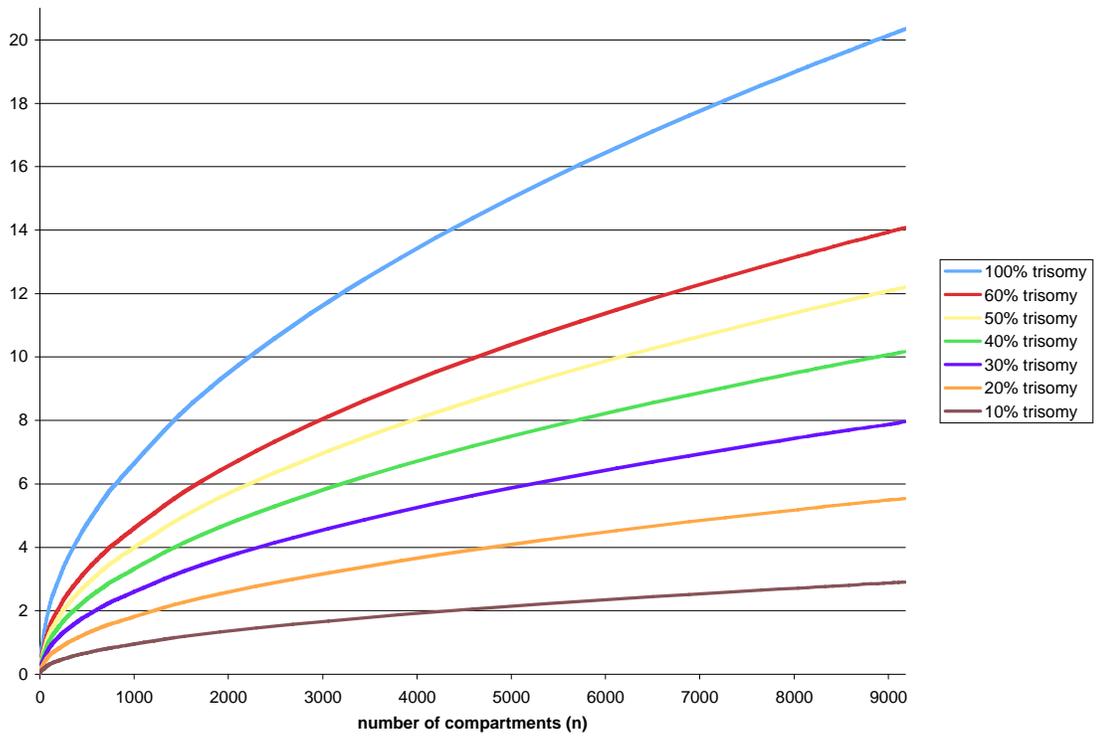

Figure 3